# Debye relaxation and 250 K anomaly in glass forming monohydroxy alcohols

S. Bauer,[1] K. Burlafinger,[2] C. Gainaru,[1] P. Lunkenheimer,[2] W. Hiller,[3] A. Loidl,[2] R. Böhmer[1]

[1] *Fakultät für Physik, Technische Universität Dortmund, 44221 Dortmund, Germany*

[2] *Experimentalphysik V, Universität Augsburg, 86135 Augsburg, Germany*

[3] *Fakultät für Chemie, Technische Universität Dortmund, 44221 Dortmund, Germany*

A previous dielectric, near-infrared (NIR), and nuclear magnetic resonance study on the hydrogen-bonded liquid 2-ethyl-1-hexanol [C. Gainaru *et al.*, Phys. Rev. Lett. **107**, 118304 (2011)] revealed anomalous behavior in various static quantities near 250 K. To check whether corresponding observations can be made for other monohydroxy alcohols as well, these experimental methods were applied to such substances with 5, 6, 7, 8, and 10 carbon atoms in their molecular backbone. All studied liquids exhibit a change of behavior near 250 K which is tentatively ascribed to effects of hydrogen bond cooperativity. By analyzing the NIR band intensities, a linear cluster size is derived that agrees with estimates from dielectric spectroscopy. All studied alcohols, except 4-methyl-3-heptanol, display a dominant Debye-like peak. Furthermore, neat 2-ethyl-1-butanol exhibits a well resolved structural relaxation in its dielectric loss spectrum which so far has only been observed for diluted monohydroxy alcohols.

# I. INTRODUCTION

Monohydroxy alcohols are often studied with the goal to learn more about the impact of hydrogen bonds on the structure and dynamics of liquids.[1,2,3,4,5] One of the consequences of their molecular association is the occurrence a very strong dielectric loss peak. On a phenomenological level, this loss peak looks similar to that found for water for which, due to its large intensity, it is exploited in the microwave oven. The dominant loss peak in water as well as in monohydroxy alcohols is typically of the Debye type, corresponding to single-exponential relaxation.[6] The molecular origin of the underlying orientational polarization has remained elusive for decades, probably because it is commonly believed that only dielectric techniques are sensitive to the Debye process. However, recently a number of non-dielectric techniques were applied to explore the nature of this relaxation in monohydroxy alcohols, among them calorimetric,[7] shear mechanical,[8] light scattering,[5] and magnetic resonance techniques.[9,10]

It was occasionally emphasized that the dielectric loss as well as the orientational polarization of the monohydroxy alcohols are much stronger than expected on the basis of their molecular dipole moment.[11] In dielectric spectroscopy this apparently enhanced relaxation strength is often described in terms of phenomenological factors, such as the Kirkwood correlation factor $g_K$ which can be calculated if the mutual molecular arrangement is known.[12] Furthermore, the dielectric strength in monohydroxy alcohols can strongly vary, and sometimes non-monotonically so, with changes in temperature,[13] in pressure,[2,14] and in concentration if these liquids are diluted by suitable solvents.[15,16]

In a recent study,[17] the dielectric strength $\Delta\varepsilon$ of 2-ethyl-1-hexanol (2E1H, $C_8H_{18}O$) was found to exhibit a peculiar change in its temperature dependence near 250 K, i.e., about 50 K above its melting point. At $T < 250$ K, $\Delta\varepsilon(T)$ varied significantly less with temperature than for $T > 250$ K. For 2E1H, anomalous trends were also found in other physical quantities: These include the absorbance $A(T)$ arising from the stretching vibration of the alcohol's OH group that can be detected in the near-infrared (NIR) spectral range at wavelengths between 1400 and 1700 nm.[17] Another property that was measured in the cited study is the chemical shift $\delta(T)$ of the hydroxyl proton which is accessible using high-resolution nuclear magnetic resonance (NMR) techniques. Most interestingly, all three quantities $\Delta\varepsilon(T)$, $A(T)$, and $\delta(T)$, each of which reflects static rather than dynamic properties, revealed the same temperature trend, including the observation of an "anomaly" near 250 K. It is tempting to rationalize these observations, e.g., in terms of peculiarities of the intermolecular structure of 2E1H or as an effect of the so-called hydrogen bond cooperativity which refers to the strengthening of these bonds when the number of molecules being part of an existing H-bond network is increasing.[18] However, before exploring further possible explanations, it is necessary to clarify whether similar observations can be made for other monohydroxy alcohols.

In particular, it would be useful to know whether the existence of this anomaly, or the temperature at which it occurs, depends on the molecular structure. Hence, we investigated monohydroxy alcohols for which the number $n$ of carbons and the position of the OH group within the molecule vary. For the present study we have chosen molecules with $n = 5$ (3-methyl-2-butanol, 3M2B), $n = 6$ (2-ethyl-1-butanol, 2E1B), $n = 7$ (5-methyl-2-hexanol, 5M2H), $n = 8$ (4-methyl-3-heptanol, 4M3H), and $n = 10$ (3,7-dimethyl-1-octanol, 3,7D1O). The properties of 2E1H, carrying an OH group in a terminal position, can be compared with those of the structural isomer 4M3H where the hydroxyl group is in the "middle" of the molecule. Due to the steric hindrance of the polar group, the central OH position in the latter case can disfavor the formation of extended hydrogen bonded structures.[19] It should be noted that most of the listed monohydroxy alcohols were already studied using dielectric spectroscopy in the low-frequency range (typically below 1 MHz) as well as by other techniques.[8,11,13,20,21,22] In particular, in Ref. 22 the glass transition temperatures determined by calorimetry ($T_{g,\text{cal}}$) and determined from the dielectrically detected structural ($\alpha$-) relaxation ($T_{g,\alpha}$) were given, e.g., for 2E1B ($T_{g,\text{cal}} = 131.0$ K, $T_{g,\alpha} = 129.4$ K), 5M2H ($T_{g,\text{cal}} = 152.1$ K, $T_{g,\alpha} = 149.8$ K), and 2E1H ($T_{g,\text{cal}} = 145.9$ K, $T_{g,\alpha} = 144.0$ K).

Our previous study on 2E1H (Ref. 17) demonstrated that the anomalous behavior in $\Delta\varepsilon(T)$, $A(T)$, and $\delta(T)$ shows up at a temperature around 250 K at which the relaxation rate of the Debye-process is larger than 1 MHz. Consequently, measurements extending up to the GHz range are necessary which we have undertaken for the present study. In most cases, we are thus able to cover more than 14 decades in frequency. In addition to performing high-frequency dielectric spectroscopy, we also recorded NIR vibrational spectra as well as $^1$H chemical shift NMR spectra for most of the monohydroxy alcohols mentioned above.

## II. EXPERIMENTAL DETAILS

The investigated monohydroxy alcohols were either purchased from Sigma Aldrich with stated purities of 98% (5M2H), 99% (4M3H, mixture of erythro and threo isomers, used for the NIR measurements), and 99% (3,7D1O) or from Alpha Aesar with stated purities of 98% (3M2B), 99% (2E1B), and 99% (4M3H, used for the dielectric measurements). All chemicals were used without further treatment except for 3,7D1O which was also measured after drying it overnight with a molecular sieve (pore size: 4 Å), prior to the NIR experiments (see Sec. III. B).

Several experimental techniques were combined to arrive at broadband dielectric spectra covering a frequency range of up to $10^{-4}$ Hz - 30 GHz. In the low-frequency range, $\nu <$ 3 MHz, a frequency-response analyzer (Novocontrol α-analyzer) was used. For the radio-frequency and microwave range (1 MHz $< \nu <$ 3 GHz), we applied a reflectometric technique, with the sample capacitor mounted at the end of a coaxial line.[23] These measurements were performed with an Agilent E4991A impedance analyzer. For the above-mentioned methods, the sample material was filled into parallel-plate capacitors, using glass-fiber spacers of 50 or 100 µm diameter to separate the capacitor plates. Part of the results at the highest frequencies of 100 MHz - 30 GHz was obtained by measuring the reflection coefficient of an open-ended coaxial line, which was directly immersed into the sample liquid, kept in a glass tube.[24,25] For this purpose, the Agilent "Dielectric Probe Kit" and an Agilent E8363B Network Analyzer were used. In addition, in the same region of 100 MHz - 30 GHz coaxial transmission measurements using a Hewlett-Packard 8510 network analyzer were carried out. For these measurements, the sample material was filled into a specially designed coaxial line, sealed with Teflon discs. The method is based on that developed by Nicholson and Ross for time domain measurements of dielectric materials.[26] This method was employed for regions with low loss for which the resolution of the open-end coaxial method was insufficient. For cooling and heating of the samples, a nitrogen gas heating system (Novocontrol Quatro), a closed cycle refrigerator (CTI-Cryogenics), and several home-made heating devices were used.

NIR spectroscopy was carried out using a CARY 2300 UV/VIS-NIR photospectrometer from Varian. All NIR spectra were recorded from 1000 nm to 2000 nm, corresponding to wavenumbers ranging from 10000 cm$^{-1}$ to 5000 cm$^{-1}$, with a wavelength resolution of 0.8 nm. The spectra were corrected by subtraction of the absorbance $A = \varepsilon(\lambda)cd$ (Lambert-Beer's law) of an empty Hellma cuvette (110 QX-2mm) and that of the windows of the vacuum chamber.[27] Here, $\varepsilon(\lambda)$ is the wavelength dependent absorption coefficient, $c$ the molar concentration of the sample, and $d$ = 2 mm the path length within the cuvette. The samples were cooled down with a closed cycle refrigerator from CTI-Cryogenics in steps of 5, 10 or 15 K and the temperature was controlled with a LakeShore 340 Temperature Controller. After ten minutes, the temperature was stable within 30 mK. To account for temperature dependent density variations, the spectra were scaled to the absorbance of the $3\nu_a(CH_3)$ band (corresponding to the second overtone of the asymmetric stretching vibration of the $CH_3$ group) located near 1190 nm at 300 K.[28]

NMR measurements of the $^1H$ isotropic chemical shift $\delta(T)$ were performed using a Bruker Avance DPX 300 spectrometer at a Larmor frequency of 300 MHz. The temperature was controlled with a BVT 3300 temperature controller from Bruker and was stable within ±0.5 K. All samples were filled into NMR tubes with an outer diameter of 5 mm. The temperature dependent spectra of the monohydroxy alcohols were scaled to their $CH_3$ resonance appearing at about 1 ppm at 298.15 K.

### III. RESULTS AND ANALYSIS
#### A. Dielectric spectroscopy

In Fig. 1(a) and (b) we show the real part and the imaginary part, respectively, of the dielectric constant for 2E1B. The real part exhibits a large relaxation strength and, upon closer inspection, a two-step behavior is revealed as a function of frequency. For low temperatures the overall relaxation strength decreases only slightly whereas above about 250 K a much more pronounced dependence shows up. At the highest temperatures it becomes increasingly difficult to resolve two steps in $\varepsilon'(\nu)$. Corresponding observations can be made from the dielectric loss. Here, in the double-logarithmic representation of Fig. 1(b), the change in intensity evolution taking place near 250 K is even more evident. For $T < 250$ K the dominant process is clearly of Debye type. Its high-frequency flank can be traced down to intensities which are about two decades lower than at the peak. Then, towards still higher frequencies two additional processes follow which, in accord with previous measurements on other monohydroxy alcohols,[7,29,30] can be identified with the α- and the Johari-Goldstein β-relaxation.[31] At sufficiently low temperatures, an α-peak maximum can clearly be resolved. This behavior is highly unusual for pure monohydroxy alcohols, as becomes obvious, e.g., by a comparison with the data on several other members of this class of glass formers, shown in Fig. 2. A spectrally well-resolved α-peak, as revealed here for neat 2E1B, was, as far as we know, heretofore only reported for significantly *diluted* monohydroxy alcohols.[27,32,33]

For a quantitative analysis of the current data we use a superposition of three relaxation processes, representing the Debye relaxation as well as the α- and the β-relaxation. For the former, a Debye function was naturally assumed. For the α-relaxation a Havriliak-Negami form (width parameters $\gamma_\alpha$ and $\alpha_\alpha$) and for the β-relaxation a Cole-Cole form (width parameter $\alpha_\beta$) was used,[34] so that overall we arrive at

$$\varepsilon^*(\nu) = \varepsilon_\infty + \frac{\Delta\varepsilon_D}{1+2\pi i \nu \tau_D} + \frac{\Delta\varepsilon_\alpha}{[1+(2\pi i \nu \tau_\alpha)^{\alpha_\alpha}]^{\gamma_\alpha}} + \frac{\Delta\varepsilon_\beta}{1+(2\pi i \nu \tau_\beta)^{\alpha_\beta}}. \quad (1)$$

Here $\varepsilon_\infty$ is the high-frequency dielectric constant. $\Delta\varepsilon_{D,\alpha,\beta}$ and $\tau_{D,\alpha,\beta}$ denote the relaxation strength and the relaxation time, respectively, of the corresponding processes. The solid lines in Fig. 1 are least-square fits using Eq. (1), simultaneously applied to real and imaginary part, and are seen to match the data excellently. For all temperatures we find that the main peak can be well fitted by the first part of Eq. (1) and, hence, for 2E1B the appearance of a dominant Debye peak is fully confirmed by the fit. For the α-process, we found that for the lower temperatures $T \leq 201$ K, the α-peak cannot be described with $\alpha_\alpha = 1$, which would correspond to a Cole-Davidson function, known to account for the α-relaxation of various glass formers.[35,36] Instead, it was necessary to use the often-employed Havriliak-Negami function with the additional parameter $\alpha_\alpha$, in order to achieve a reasonable fit of the transition region between Debye- and α-peak. In most cases we found $\gamma_\alpha$ to be close to one, implying that the α-process is describable by a nearly symmetric spectral shape. Symmetrically broadened spectra were already clearly documented for binary mixtures involving alcohols.[27,32] Due to the overlap of the three peaks, the uncertainties of the width parameters obtained from the fits are rather large and clear statements on their temperature development cannot be made. Other fit parameters and their temperature dependences will be presented and discussed in Sec. IV C, below. We just note here that the glass transition temperatures $T_{g,\alpha}$, determined on the basis of our measurements, are in good agreement with those for which literature data are available.[22] For 3M2B and 3,7DM1O we find $T_{g,\alpha} = 137$ K and $T_{g,\alpha} = 153$ K, respectively.

In Fig. 2(a), (b), and (c) we present the dielectric losses $\varepsilon''(\nu)$ for 3M2B, 5M2H, and 3,7D1O, respectively. Overall, these liquids reveal the same general trends as 2E1B. In particular, a change in the temperature dependence of $\Delta\varepsilon_D$ is always observed near 250 K in accord with the observations made for 2E1B (cf. Fig. 1) and with our previous report on 2E1H.[17] In 3M2B, 5M2H, and 3,7D1O the α-process is revealed only by its high-frequency flank; its low-frequency flank is completely submerged under the Debye peak. Hence, while the experimental data could be fitted using a Cole-Davidson function for the α-relaxation (i.e., $\alpha_\alpha = 1$ in Eq. (1), an unambiguous statement regarding the curve form of the α-peak is not possible for these liquids. In any case, the main peak could be well fitted using a Debye function, again demonstrating the presence of a dominating monodispersive relaxation process in these monohydroxy alcohols. An inspection of Fig. 1(b) and Fig. 2 shows that the strengths of the β process, relative to $\Delta\varepsilon_D$, as well as their temperature dependences differ. For 2E1B the β process is very weak and exhibits little temperature dependence, while for 5M2H the strength of the secondary relaxation increases strongly with increasing temperature.

The data in Fig. 1 and Fig. 2 as well as previously published results[17,37] suggest that the anomaly in $\varepsilon'$ and $\varepsilon''$ seen near 250 K is a universal feature of monohydroxy alcohols. Therefore, we also studied the monoalcohol 4M3H, for which, in contrast to what is seen in Fig. 1 and Fig. 2, the total dielectric strength is known to be very small.[13] For 4M3H the real and imaginary parts of $\varepsilon^*$ are shown in Fig. 3 and a large difference to the alcohol data presented in Fig. 1 and Fig. 2 is immediately obvious. With a static dielectric constant $\varepsilon_s < 3.5$, the total dielectric strength of 4M3H is not only much smaller as compared to the other alcohols, but also displays little temperature dependence for $T < 250$ K, see Fig. 3(a). However, for $T > 250$ K $\Delta\varepsilon$ strongly increases with increasing temperature, in obvious contrast to what is found for the other alcohols. While from $\varepsilon'(\nu)$ alone it is hard to decide without further analysis whether more than one relaxation process is present in 4M3H, Fig. 3(b) demonstrates that this is possible from $\varepsilon''(\nu)$. A two-peak structure is revealed to exist for temperatures from ~170 to ~220 K. Curiously, the strength of the Debye process is weaker than that of the α-relaxation. Due to differences in the temperature dependences of $\tau_D$ and $\tau_\alpha$, above about 250 K only a single peak is resolved which becomes rather intense upon heating. The lines in Fig. 3 are again fits using Eq. (1), but before discussing the resulting parameters in Sec. IV C, we will first present results obtained using NIR and NMR spectroscopy in Sec. III B and in Sec. III C, respectively.

## B. Temperature dependent near-infrared spectroscopy

### *1. Spectra and peak intensities*

In Fig. 4 NIR spectra are presented for all investigated alcohols. The spectra were measured at 300 K, ranging from $\lambda = 1320$ to 1700 nm, and they were corrected as detailed in Sec. II. The inset in Fig. 4 illustrates the assignment of various OH overtone bands (called α-, β-, γ-, and δ-band) in terms of the bonding state of the alcohol's hydroxyl group. The band around $\lambda_{\alpha/\beta} = 1410$ nm is assigned to the first overtone of free OH vibrations (α-state) which is close to the vibrational wavelength of the proton accepting OH groups (β-state). In the following, the absorbance at the peak of this α/β-band is labeled $A(\lambda_{\alpha/\beta})$ and marked by a red arrow in Fig. 4. The small feature around $\lambda_\gamma = 1430$ nm is due to the absorbance of proton donating OH groups [γ-state, $A(\lambda_\gamma)$], whereas the broad band near $\lambda_\delta = 1600$ nm reflects vibrations of strongly hydrogen bonded OH groups [δ-state, $A(\lambda_\delta)$].[27,38] The wavelength region between 1430 and 1600 nm can be assigned to other species like (cyclic) multi-mers or branched aggregates.[39,40]

In Fig. 4 the absorbance spectra of the different alcohols are displayed from top to bottom according to increasing $A(\lambda_{\alpha/\beta})$. Since the α/β-band of 3,7D1O displays the smallest intensity and 4M3H the largest one, the degree of hydrogen bonding decreases from 4M3H to 3,7D1O which should also be reflected by the δ-band. Rather than relying on $A(\lambda_\delta)$ alone, we found that the ratio $\rho = A(\lambda_\delta)/A(\lambda_{\alpha/\beta})$ of the absorbances of the δ- and the α/β-bands is a good indicator for the degree of hydrogen bonding. On the other hand, the position of the δ-band displays a systematic red shift when going from bottom to top in Fig. 4. This red shift means that in systems with stronger hydrogen bonds the covalent OH bond is slightly weakened and the corresponding stretching frequency becomes smaller.

It is not completely straightforward to extract quantitative results from $\rho$ as the α/β-band strongly overlaps with a CH combination band.[41,42] Assuming that the different monohydroxy alcohols contribute all similarly to this CH combination band, one can nevertheless use $\rho$ to compare the degree of hydrogen bonding in the various monohydroxy alcohols. For 3,7D1O, 2E1H, and 2E1B with the OH group located at a terminal position, the so called 1-alkanols, $\rho$ is close to one: 1.03 for 3,7D1O, 0.99 for 2E1H, and 1.09 for 2E1B. For the 3-alkanol 4M3H, with $\rho = 0.52$, the absorbance ratio is quite different. Here, a pronounced absorbance is seen around 1480 nm that presumably stems from cyclic dimers.[39,40] This feature is much stronger than the band due to strongly hydrogen bonded OH groups in linear chains.[9] The 2-alkanols 5M2H and 3M2B exhibit both an intermediate absorbance ratio, $\rho = 0.90$. These numbers imply that the position of the OH group within the alcohol molecule has a strong impact on the degree of hydrogen bonding. The steric hindrance experienced by the OH group is thus much smaller for the 1-alkanols 3,7D1O, 2E1H, and 2E1B and smaller for the 2-alkanols 5M2H and 3M2B as compared to the 3-alkanol 4M3H.

In Fig. 5 NIR spectra of 3,7D1O, 5M2H, and 4M3H are presented covering the temperature range from 300 K down to about 135 K. For 2E1H, corresponding spectra were already published.[17] Quantitatively, all these monohydroxy alcohols show the same temperature dependence: With decreasing temperature, the absorbances of the α/β- and of the γ-bands are decreasing and concomitantly the absorbance of the δ-band is strongly increasing.

In Fig. 6 the peak absorbances $A(\lambda_\delta)$, $A(\lambda_\gamma)$, and $A(\lambda_{\alpha/\beta})$ of the spectra are plotted versus temperature. For 3,7D1O, see Fig. 6(a), the absorbance of the δ-band increases linearly with decreasing temperature down to ~150 K and saturates for lower temperatures. This is in good agreement with the glass transition temperature $T_{g,\alpha} = 152.8$ K extracted from dielectric measurements, cf. Sec. III A. Thus, the number of strongly hydrogen bonded OH oscillators is increasing down to ~150 K and then remains constant. The absorbances of the α/β- and of the γ-bands decrease strongly with decreasing temperature and show a much reduced $T$ dependence below ~250 K. Overall, 5M2H, see Fig. 6(b), and 4M3H, see Fig. 6(c), exhibit similar temperature dependences of $A(\lambda_{\alpha/\beta})$ and $A(\lambda_\gamma)$ compared to those of 3,7D1O. For the δ-bands small additional changes of slope appear as highlighted by the straight lines in Fig. 6. A saturation of $A(\lambda_\delta)$ is reached near 150 K for 5M2H and near 155 K for 4M3H.

*2. Wavelengths of maximum absorbance*

For several monohydroxy alcohols we also analyzed the temperature dependences of the wavelengths read out near 1600 and near 1410 nm, i.e., at the well resolved absorbance peaks. With decreasing temperature, the δ-band of 3,7D1O, 2E1H, 5M2H, and 4M3H shows a strong red shift of approximately 25 nm, see Fig. 7(a).[43] In accord with previous IR investigations this red shift can be interpreted to be due to an increasing degree of hydrogen bond cooperativity:[18,44,45,46] This term means that the formation of a hydrogen bridge enhances the electron density at the lone pairs of the proton donating oxygen atom and therefore increases the ability for proton acceptance.[47] The change in electron density is in turn accompanied by a stretching of the covalent OH bond. Approximating the OH group as a harmonic oscillator with a vibrational frequency $\omega = \sqrt{k/m_r}$, an increase of the covalent OH bond length results in a decrease of the force constant $k$ and of $\omega$ since the reduced mass $m_r$ remains constant. The red shift of the δ-band observed for decreasing temperature and hence an increased hydrogen bond cooperativity suggests that more and more alcohol molecules are added to existing clusters upon cooling. These hints are in harmony with the arguments that will be presented in Sec. IV.A.

For the studied monohydroxy alcohols also the wavelengths of the α/β-band exhibit an apparent red shift when cooling down to about 240 K and then they saturate, see Fig. 7(b). In order to rationalize the apparent temperature dependence of $\lambda_{\alpha/\beta}$ it is important to recall that the absorbances arising from the α-state, appearing at a wavelength $\lambda_\alpha$, and the β-state, appearing at $\lambda_\beta$, are overlapping. With decreasing temperature the number of free, non-bonded alcohol molecules, and hence $A(\lambda_\alpha)$ is expected to decrease at the expense of $A(\lambda_\beta)$. This change of population will hence appear as a shift of an ensemble averaged wavelength $\lambda_{\alpha/\beta}$. This argument is in line with the large degree of similarity of the wavelengths of the two bands that could be resolved by two-dimensional NIR experiments on 1-octanol:[39] It was found that $\lambda_\alpha$ = 1406 nm and $\lambda_\beta$ = 1412 nm.

Any "shift" in $\lambda_{\alpha/\beta}$ that we detect experimentally can thus be regarded as a lower limit of $\Delta\lambda_{\alpha/\beta} < \lambda_\beta - \lambda_\alpha$.[48] In the available temperature range we find that $\Delta\lambda_{\alpha/\beta} = \lambda_{\alpha/\beta}(140\text{ K}) - \lambda_{\alpha/\beta}(300\text{ K})$ = 4 nm for 3,7D1O, 2.5 nm for 2E1H, and 2 nm for 5M2H while for 4M3H $\Delta\lambda_{\alpha/\beta}$ is only 1 nm. These numbers indicate that in the monohydroxy alcohols with a large Debye process $\Delta\lambda_{\alpha/\beta}$ is significant. Only in 4M3H, the change in the population of the α- and β-state is small: The absorbance of the β-state is visible as a small shoulder of the α/β-band at its high-wavelength flank, see Fig. 5(c), and the equilibrium of this α/β-population does not vary much upon cooling.

### C. Proton nuclear magnetic resonance

The $^1$H isotropic chemical shift $\delta$, which is a metric of the shielding of the external magnetic field by the local electronic environment of a given proton, was measured at room temperature for 3,7D1O, 2E1H, 5M2H, and 4M3H. For the hydroxyl proton of all alcohols we find $\delta_{OH}$ = (5.30 ± 0.05) ppm except for 4M3H with $\delta_{OH}$ = 4.6 ppm (not shown). Consequently, the hydroxyl group of 4M3H is shielded more than the OH groups of the other monohydroxy alcohols indicating a smaller degree of hydrogen bonding. These findings are in line with the absorbance ratio $\rho$ defined in Sec. III B.

For two monohydroxy alcohols (2E1H, Ref. 17, and 3,7D1O) also temperature dependent data, $\delta(T)$, were recorded. In Fig. 8 we show the $^1$H isotropic chemical shift spectra for 3,7D1O which were measured down to temperatures at which the hydroxyl line starts to broaden significantly. Most of the $CH_2$ and $CH_3$ resonances of this alcohol are found between 1 and 2 ppm except the resonance for the α-$CH_2$ group, the methylene next to the OH group within the molecule, which appears at around 3.8 ppm. The most so-called low-field, i.e., the most de-shielded signal is due to the hydroxyl proton.

The $^1$H resonances of free and hydrogen bonded OH species can not be resolved spectroscopically from each other since they merge due to the effect of motional narrowing: While the local magnetic field experienced by a hydroxyl proton is affected differently if located in bonded (b) or in non-bonded (n) OH groups (which may associated with shifts of $\delta_b$ and of $\delta_n$, respectively, see also Refs. 49 and 50), the time scale $\tau$ of this H bond switching is much smaller than the inverse spectral separation $|\delta_b - \delta_n|^{-1}$. Under these circumstances only a single, narrowed resonance line can be observed in an NMR spectrum.[17]

The chemical shift of the OH group $\delta_{OH}(T)$ for 3,7D1O moves towards higher frequency with decreasing temperature whereas the positions of the resonance lines of $CH_2$ and $CH_3$ groups are temperature independent, see Fig. 8. The shift of $\delta_{OH}(T)$ with decreasing temperature results from the fortification of hydrogen bonds because the shielding of the hydroxyl proton is reduced as the H-bonded oxygen draws more electron density away from the proton.[51] Importantly, $\delta_{OH}(T)$ displays a non-linear temperature dependence with a kink appearing near 250 K. This feature will be discussed together with results for the other alcohols in Sec. IV B.

## IV. DISCUSSION
### A. Average cluster size of alcohol aggregates

From the peak absorbances $A(\lambda_\gamma)$ and $A(\lambda_\delta)$, one can extract the temperature dependence of the average number $\langle n \rangle$ of alcohol molecules in a cluster. In a first step we assume that these aggregates are not branched and do not form rings but we will come back to these scenarios later. Let us therefore consider that $N$ end-to-end chains are present in the sample. Trivially, each chain is terminated by one proton accepting and by one proton donating terminal hydroxyl group. Hence, the total number B of these acceptors and the total number Γ of the donors both are equal to $N$ (i.e., $N = B = Γ$). Therefore, if the total number Δ of strongly hydrogen bonded OH groups *within* all of the $N$ end-to-end chains is known, the average number of OH groups (and for the present case of *mono*hydroxy alcohols thus of molecules) forming an end-to-end chain is given by

$$\langle n \rangle = \frac{\Delta}{\Gamma} + 2. \qquad (2)$$

Here the term "+2" represents the two terminal OH groups in each end-to-end chain. Assuming that the measured peak intensities of the γ- and δ-bands are proportional to $\Gamma$ and $\Delta$ (with the proportionality factors given by the oscillator strengths $I_\Gamma$ and $I_\Delta$, respectively), an average cluster size of the end-to-end chains can thus be deduced. However, the absorbance $A(\lambda_\gamma)$ does not solely reflect $\Gamma$, but also contains a $T$-independent contribution $K_\Gamma$ from a CH combination band.[41,42] Hence, one may write $\Gamma \cdot I_\Gamma = A(\lambda_\gamma) - K_\Gamma$. Similar arguments for the δ-band yield $\Delta \cdot I_\Delta = A(\lambda_\delta) - K_\Delta$. Here $K_\Delta$ again denotes background contributions, which at least for 2E1H, can be estimated from NIR spectra of homologous OH free liquids such as 2-ethyl-1-hexylbromide ($K_\Gamma \sim 0.05$ and $K_\Delta \sim 0$).[27] With these considerations the average cluster size of linear alcohol chains $\langle n \rangle$ follows as

$$\langle n \rangle = \frac{I_\Gamma}{I_\Delta} \cdot \frac{A(\lambda_\delta) - K_\Delta}{A(\lambda_\gamma) - K_\Gamma} + 2 \qquad (3)$$

The ratio of the oscillator strengths, $I_\Gamma/I_\Delta$, appearing in Eq. (3) is not accessible from our experimental data, but in view of previous experiments[52] we assume that it is close to unity.

Consequently, we are able to estimate the average cluster size $\langle n \rangle$ of 2E1H from the NIR data presented in Ref. 17: As the number of free alcohol molecules [$\propto A(\lambda_{\alpha/\beta})$] as well as of linear chains [$\propto A(\lambda_\gamma)$] is decreasing down to 150 K and the number of strongly hydrogen bonded alcohol molecules within linear alcohol chains [$\propto A(\lambda_\delta)$] is increasing with decreasing temperature, the average cluster size has to be increasing with decreasing temperature. According to Eq. (3), $\langle n \rangle$ for 2E1H ranges from $\langle n \rangle = 4$ at 300 K up to $\langle n \rangle = 14$ at 150 K. The temperature dependences of the peak absorbances for the other monohydroxy alcohols are similar to those of 2E1H, see Fig. 6, implying a similar temperature dependence of the average cluster size.

Let us now briefly consider *branched* supramolecular structures. Each branch in an end-to-end chain contributes with an additional proton *donating* end group to $A(\lambda_\gamma)$ and concomitantly reduces the number of strongly hydrogen bonded OH groups, i.e., also $A(\lambda_\delta)$.[53] The existence of branched end-to-end chains thus decreases the calculated $\langle n \rangle$, see Eq. (3). However, from theoretical work the fraction of branched monohydroxy alcohol structures is usually found to be smaller than ~10%,[53,54,55] and hence this effect is of the same magnitude as the resulting uncertainty of $\langle n \rangle$.

Finally, there may not only be linear and branched species but also ring like alcohol structures that display neither β- nor γ-states. Depending on the number of H-bonded alcohol molecules these rings usually lead to absorption contributions at wavelengths between $\lambda_\gamma$ and $\lambda_\delta$. It has been found that rings composed of about 4 or more members contribute to $A(\lambda_\delta)$ (Ref. 56) and then would tend to increase the calculated $\langle n \rangle$.

For alcohols such as 4M3H, for which rings can be expected to be the dominant mesoscopic structure[57] so that the effective dipole moment is largely canceled and the Debye process is weak, these simple considerations are not applicable. Under these circumstances experiments need to be performed which are not only sensitive to local stretching vibrations of covalently bonded molecular groups, but also to the properties of more extended and thus more weakly (hydrogen) bonded aggregates. Weak bonding implies lower vibrational frequencies so that far infrared spectroscopy could provide a suitable means to study such aggregates like, e.g., ring-like structures.

However, for systems with a strong Debye process, rings can be expected to play a minor role and so the average cluster size $\langle n \rangle$ as calculated on the basis of Eq. (3) should provide a useful measure. Comparison can be made with previously reported dielectric data on viscous monohydroxy alcohols[9] which suggested an end-to-end chain comprising about 5-10 alcohol molecules. From molecular dynamics and Monte-Carlo simulations of various n-alcohols in their more fluid state clusters involving 3...7 molecules were deduced.[54,55] All these results are in good qualitative agreement with $\langle n \rangle = 4...14$ as estimated above on the basis of NIR data. Hence, from the present analysis of the various absorbance bands and by making plausible assumptions regarding the cluster topology we obtained mean cluster sizes in agreement with previous estimates.

**B. The 250 K anomaly**

From dielectric, NIR, and NMR spectroscopy of several alcohols we found indications for peculiar behaviors showing up near 250 K. In order to discuss these observations in a more coherent way, in Fig. 9 we plot the temperature dependence of the absorbance of the α/β-state $A(\lambda_{\alpha/\beta})$, the inverse dielectric relaxation strength $1/\Delta\varepsilon(T)$, as well as the inverse chemical shift $1/\delta_{OH}(T)$ of the hydroxyl group in a scaled representation. It is seen that, whenever available, these quantities agree with each other for a given substance. This agreement confirms the conclusions previously drawn for 2E1H.[17] The temperature evolution of the hydrogen bond population with its particular behavior near 250 K is now also demonstrated for other glass formers on the basis of three independent, essentially static quantities. This finding calls for structural studies to better understand the origin of the anomaly near 250 K, e.g., in terms of possible bonding specificities.

For the current discussion let us now focus on the absorbance results. The temperature dependence of the NIR absorbance bands can be traced back to a change in the population of the participating vibrational states. Let us first assume that only the populations of *two* relevant states co-exist in equilibrium.[58] Identifying these with non hydrogen bonded [∝ $A(\lambda_{\alpha/\beta})$] and strongly hydrogen bonded OH species [∝ $A(\lambda_\delta)$], one can use the van't Hoff equation

$$\log_{10} K \propto \log_{10}\frac{A(\lambda_{\alpha/\beta})}{A(\lambda_\delta)} = -\log_{10}\rho = -\frac{\Delta H}{R\ln(10)}\cdot\frac{1}{T} + \text{const.} \qquad (4)$$

that relates the enthalpy change $\Delta H$ to the experimentally accessible equilibrium "constant" $K$. R is the gas constant. From a plot of $\log_{10}\rho^{-1}$ versus the inverse temperature, see Fig. 10, one recognizes two distinct regimes separated by an intermediate region between 250 and ~210 K (open symbols in Fig. 10). Fitting the data with Eq. (4) one can extract the enthalpy changes $\Delta H$ below and above these temperatures, see Table I. For $T > 250$ K the enthalpies for the 1-alkanols 3,7D1O and 2E1H are about 9 kJ/mol and agree within experimental error. Somewhat larger $\Delta H$ are found for 5M2H and 4M3H which is attributed to the steric hindrance experienced by the OH group. These enthalpies are comparable with those for other hydrogen bonded systems: Worley and Klotz[58] found 9.9 kJ/mol for HOD and Barkatt and Angell[59] report 10.9 kJ/mol for sorbitol.

For temperatures below about 210 K, the enthalpy yielded by our simple data analysis is only 1-2 kJ/mol. This $\Delta H$ is evidently much smaller than required for the formation of hydrogen bonds. Hence, at least in the $T$ range below 210 K the simple assumption of a two-state scenario involving just open and closed hydrogen bonds, contributing to $A(\lambda_{\alpha/\beta})$ and to $A(\lambda_\delta)$, respectively, breaks down. A glance at Fig. 5 shows that between $\lambda_{\alpha/\beta}$ and $\lambda_\delta$, at which presumably small ring-type multi-mers contribute, the spectra undergo significant changes with temperature. Unfortunately, these are hard to quantify unambiguously. Neglecting the minor temperature dependence of $A(\lambda_{\alpha/\beta})$ for $T < 210$ K (cf. Fig. 6) one can nevertheless conclude that changes along the ordinate axis in Fig. 10 reflect changes essentially in $A(\lambda_\delta)$. Consequently, the empirically determined enthalpy changes of $\leq 2$ kJ/mol hint at the prevalence of a temperature dependent population equilibrium involving chain-like and presumably ring-type supramolecular hydrogen bonded structures that are energetically almost equivalent.

To summarize this subsection, we conclude that at high temperatures ($T > 250$ K) the change in the hydrogen bond population is essentially driven by the equilibrium between open and closed bonds, while at lower temperatures a redistribution among different (e.g., chain- and ring-type) supramolecular hydrogen bonded structures prevails.

**C. Relaxation times, fragility, and decoupling**

From fits to our dielectric data as described in Sec. III A we determined the characteristic time scales corresponding to the Debye process, to the structural relaxation, as well as to the Johari-Goldstein secondary relaxation. In order to compare the time constants of all presently studied substances, they are shown in an Arrhenius plot in Fig. 11. The results for the $\alpha$-relaxation times were fitted using a Vogel-Fulcher law

$$\tau = \tau_0 \exp\left(\frac{D}{T-T_0}\right). \tag{5}$$

Here $\tau_0$ is a pre-exponential factor and $D$ as well as $T_0$ are empirical constants. The fragility index $m$,[60] defined as the slope of the $\tau(T)$ curve in the representation of Fig. 11, shows some variation among the different alcohols. We find that there is a general trend that $m$ increases with the number of carbons in the molecular backbone from ~35 for 3M2B to ~75 for 3,7D1O (cf. Fig. 11) which is similar to the range identified previously for another set of monohydroxy alcohols.[22] Generally, in the presence of a sufficiently strong Debye process which overlays the structural relaxation, the determination of fragility indices is not very precise. Likewise, its determination is also impeded for substances like 4M3H for which Debye and structural relaxation times are separated by a factor of less than 10.

The β-relaxation process exhibits a large degree of similarity for all studied alcohols As observed in Fig. 11, a thermally activated behavior is obeyed near and below the glass transition of the various liquids. In terms of Eq. (5) this means that $T_0 = 0$ and that $D$ is identified with an activation energy $E_\beta$. From the current data we find $E_\beta = (3600 \pm 400)$ K [equivalent to $(30 \pm 3)$ kJ/mol] which is quite typical for molecular glass forming systems.[61]

Let us now turn to the Debye relaxation times $\tau_D$ which in Fig. 11 were all shifted upwards by 3 decades for visual clarity. Fig. 11 documents in an unambiguous fashion that the decoupling of $\tau_D$ from the structural relaxation varies vastly among the different alcohols. This decoupling is seen even more clearly in Fig. 12 where we plotted $\log_{10}(\tau_D/\tau_\alpha)$ as a function of $\log_{10}\tau_\alpha$. Overall it is evident that the decoupling of time scales is strongly temperature dependent and for each substance it is maximum when the structural relaxation time is about 0.1 ms. Findings similar to the latter were presented and rationalized for various other alcohol systems before.[10,11,37] From Fig. 12 one recognizes that for 2E1B the decoupling ratio is up to about $10^4$. While such large decoupling ratios were previously found for suitably *diluted* alcohols, the one for 2E1B is larger, as far as we know, than reported for any other *neat* alcohol. Interestingly, all curves in Fig. 12 can be interpreted to approach $\tau_D/\tau_\alpha = 1$ when $\tau_\alpha$ reaches values of the order $10^{-12}$ - $10^{-13}$ s (cf. the lines in Fig. 12). This corresponds to very high temperatures,[62] at which the average cluster size of the alcohol chains can be assumed to approach one.

## V. SUMMARY

Previous measurements on the 1-alkanol 2E1H indicated that there might be an anomaly in the hydrogen bond equilibrium near 250 K. In the present work we combined NIR, dielectric, and NMR spectroscopy to further study these effects also including alcohol molecules comprising not only 8, but also 5, 6, 7, and 10 carbon atoms. Furthermore, several 2- and 3-alkanols were selected for which the OH group was located not only at the terminal site but also in other positions along the (branched) alkyl chain. The dielectric measurements were performed in a frequency range covering up to 14 decades from the glassy state up to temperatures much above the melting point. For 2E1B, 3M2B, 5M2H, and 3,7D1O dominant Debye processes were found that exhibit different degrees of decoupling from the structural relaxation. For 2E1B the decoupling ratio was enormous (~$10^4$) so that the α relaxation peak could be better resolved in the experimental spectra than in other monohydroxy alcohols. 4M3H, the alcohol for which the hydroxyl group has the largest distance to a terminal site, exhibits not only the smallest decoupling ratio but also the weakest Debye process with an amplitude smaller than that of the α-response. These observations can be rationalized in terms of the sterical hindrance experienced by the OH group, an effect which obviously disfavors the formation of linear end-to-end structures that otherwise would lead to a large end-to-end dipole moment.

A comparison of the OH overtone stretching modes obtained from NIR spectra confirmed these sterical effects and allowed us to investigate the thermal evolution of the hydrogen bond population in a more quantitative fashion. This was achieved by considering the positions, intensities, and spectral widths of various NIR absorption bands. In particular, the intensity of the γ-band, reflecting the proton donating OH groups, was compared with that of the δ-band, which stems from the strongly bonded OH groups that can be found in non-terminal sites of extended supramolecular alcohol aggregates. The topologies of the latter can range from rings and branched structures to linear clusters. Also referring to literature data including simulation work, the first two topologies were argued to be of minor importance for all studied alcohols except for 4M3H. For dominant linear structures an NIR based estimate of the length of alcohol end-to-end chains was provided. We found that an average chain contains of about 4 molecules near room temperature and that it increases by more than a factor of three when approaching the glass transition temperature.

By comparing the NIR results with chemical shift data and with dielectric relaxation strengths, an anomalous behavior was confirmed to occur near 250 K for all studied alcohols. Rationalizing the NIR data in terms of a simple, effective two-state approach, this temperature might mark a transition from a low-temperature regime characterized by a redistribution among different, energetically almost degenerate supramolecular hydrogen bonded structures, to a high-temperature range in which these structures disintegrate more rapidly. Hence the temperature dependence in the latter range reflects the effective hydrogen bond enthalpies. This peculiar change of behavior near 250 K was thus rationalized in terms of hydrogen bond cooperativity, an effect which stabilizes supramolecular structures more and more below this temperature. It will be interesting to study the resulting changes in bonding characteristics by suitable scattering methods.

**ACKNOWLEDGMENT**

This project was financially supported by the Deutsche Forschungsgemeinschaft under Grant No. BO1301/8-2 and partly via the Research Unit FOR1394 which is highly appreciated.

**Table I:**

Enthalpy changes in kJ/mol extracted from fits using Eq. (4) to the data presented in Fig. 10.

|        | $T > 240$ K   | $T < 240$ K     |
|--------|---------------|-----------------|
| 3,7D1O | 9.1 ± 0.4     | 1.05 ± 0.05     |
| 2E1H   | 9.2 ± 0.6     | 1.10 ± 0.09     |
| 5M2H   | 11.8 ± 0.7    | 1.51 ± 0.03     |
| 4M3H   | 14.0 ± 1.0    | 1.92 ± 0.06     |

[62] In fact in this time region the so-called boson peak is located, see Ref. 35, and the material decomposes or evaporates before $\tau_\alpha \leq 10^{-12}$ s is actually reached.

**Figure Captions**

Fig. 1

(Color online) Broadband dielectric relaxation spectra of the monohydroxy alcohol 2E1B. Frames (a) and (b) show the real and the imaginary part, respectively, of the complex permittivity, demonstrating that Debye process and structural relaxation are well separated and that the Debye process is much stronger than the α-relaxation. At low temperatures a β-process is identified. The solid lines represent fits using Eq. (1). Temperatures are given in Kelvin.

Fig. 2

(Color online) Broadband dielectric loss spectra of the monohydroxy alcohols 3M2B, 5M2H, and 3,7D1O are given in frames (a), (b), and (c), respectively. For all substances the Debye process is much stronger than the α-relaxation. Secondary Johari-Goldstein relaxations are well resolved in all cases. The solid lines represent fits using Eq. (1). Temperatures are given in Kelvin.

Fig. 3

(Color online). (a) Real part and (b) imaginary part of the complex dielectric constant of 4M3H. Unlike to what is seen in Fig. 1 and Fig. 2, for 4M3H the strength of the Debye peak is *smaller* than that of the α-process. It should be noted that at temperatures $T > 250$ K the dielectric strength increases significantly. The solid lines represent fits using Eq. (1). The dashed lines show the deconvolution of the spectrum at 177 K into Debye-, α-, and β-peak. Temperatures are given in Kelvin.

Fig. 4

(Color online) (a) NIR spectra of all investigated monohydroxy alcohols in the wavelength range of the first OH overtone. The spectra were recorded at 300 K and except for 4M3H they were shifted upwards by (multiples of) 0.12 for clarity. The arrows indicate the positions of the α/β-band ($\lambda_{\alpha/\beta}$ ~ 1410 nm), γ-band ($\lambda_\gamma$ ~ 1430 nm), and δ-band ($\lambda_\delta$ ~ 1600 nm). The molecular assignment of these bands is illustrated.

Fig. 5

(Color online) Representative temperature dependent NIR spectra of (a) 3,7D1O, (b) 5M2H, and (c) 4M3H are shown ranging from 300 K down to the indicated temperatures in steps of 30 K, except for the lowest temperatures. The arrows indicate the positions of the various bands as well as the direction of absorbance change with decreasing temperature.

Fig. 6

(Color online) Peak amplitudes of the absorbances of the α/β-, γ-, and δ-band read off from the spectra presented in Fig. 5 for (a) 3,7D1O, (b) 5M2H, and (c) 4M3H. The straight lines are drawn to guide the eye and serve to emphasize the change of slope around 250 K [or near 190 K for $A(\lambda_\delta)$ of 5M2H] as well as near $T_{g,\alpha}$ (highlighted by arrows).

Fig. 7

(Color online) Temperature dependence of the wavelengths of (a) the δ- and (b) the α/β-band for four monohydroxy alcohols. The data were read out at the peak absorbances $A(\lambda_\delta)$ and $A(\lambda_{\alpha/\beta})$ from the spectra shown in Fig. 5 or in Ref. 17. The lines in frame (a) emphasize that the wavelengths $\lambda_\delta$ of all alcohols exhibit essentially the same temperature dependence. Some representative error bars are included. Note the expanded ordinate scale in frame (b). The open symbols represent wavelength for dried 3,7D1O.

Fig. 8

(Color online) Temperature dependent $^1$H isotropic chemical shift spectra for 3,7D1O. The resonance position of the hydroxyl group, $\delta_{OH}(T)$, exhibits a low-field shift with decreasing temperature whereas the CH$_2$ and CH$_3$ resonances remain unaffected. The latter resonances are only shown for the highest and for the lowest temperature.

Fig. 9

(Color online) NIR peak absorbance (circles), relaxation strength (diamonds), and $^1$H chemical shift (triangles) for (a) 3,7D1O, (b) 2E1H, (c) 4M3H, and (d) 5M2H. Dielectric data for 3M2B and for 2E1B are shown in frame (e). The scale on the right hand side reflects the inverse dielectric relaxation strength $1/\Delta\varepsilon_D$. The outer scale on the left hand side refers to the absorbance $A(\lambda_{\alpha/\beta})$ whereas the inner scale in (a) and (b) corresponds to the inverse proton chemical shift $1/\delta_{OH}$ in ppm$^{-1}$. The solid lines emphasize the roughly linear dependence found at low temperatures. All quantities and liquids display an anomalous behavior at a temperature of ~250 K which is highlighted by the dashed line.

Fig. 10

(Color online) Van't Hoff plot for four monohydroxy alcohols. Two distinct regimes are visible which are separated by an intermediate regime that is marked by open symbols. These data as well as those below 145 K were not used for determining the enthalpy using Eq. (4). The enthalpies resulting from the corresponding fits are summarized in Table 1. For clarity, the fits are depicted for 4M3H and 5M2H, only.

Fig. 11

(Color online) Dielectric relaxation times of the various alcohols are shown in an Arrhenius plot. For visual clarity the Debye relaxation times (crosses) were multiplied by a factor of 1000. The lines through the $\tau_D$ data points are guides to the eye. The full symbols show average α-relaxation times. Fits using Eq. (5) are represented as solid lines. The glass transition temperatures resulting from $T_{g,\alpha} = T(\tau_\alpha = 10^2 \text{ s})$ are listed in the figure. The straight dashed lines demonstrate that the β-relaxation times (open triangles) follow a thermally activated behavior. The data for 2E1H were taken from Ref. 17.

Fig. 12

(Color online) The logarithmic time scale decoupling ratio $\log_{10}(\tau_D/\tau_\alpha)$ is plotted versus the structural relaxation time. For each substance the decoupling is maximum when $\tau_\alpha \approx 10^{-4}$ s. An extraordinarily large maximum ratio of about $10^4$ is exhibited by 2E1B. The solid lines guide the eye.

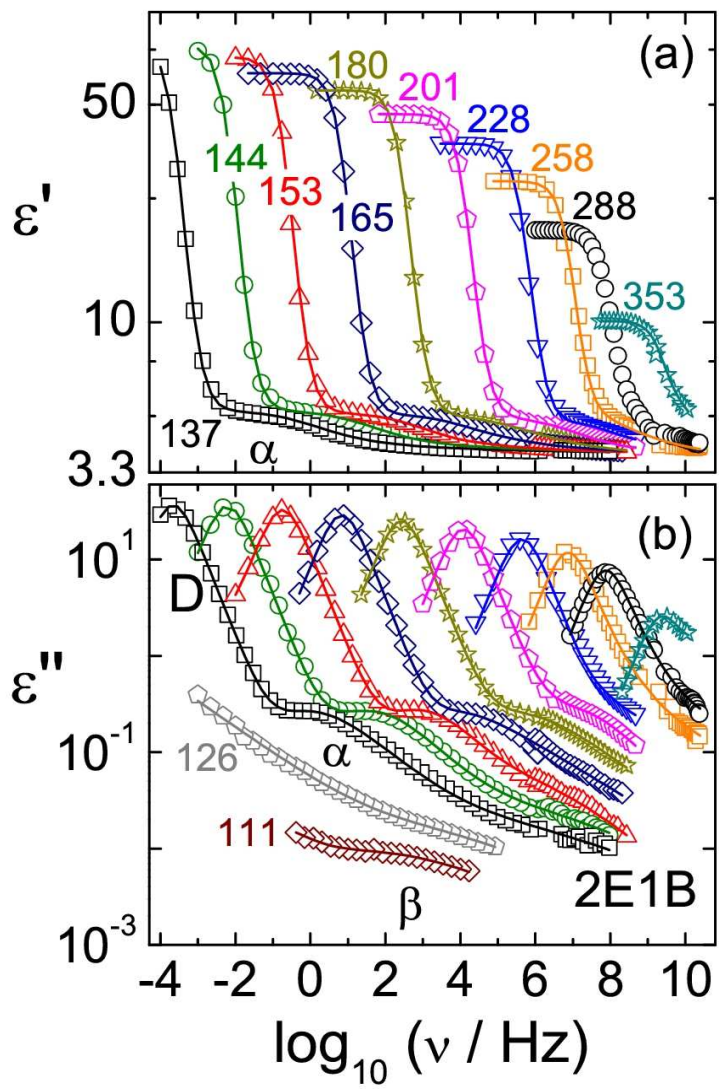

Fig. 1

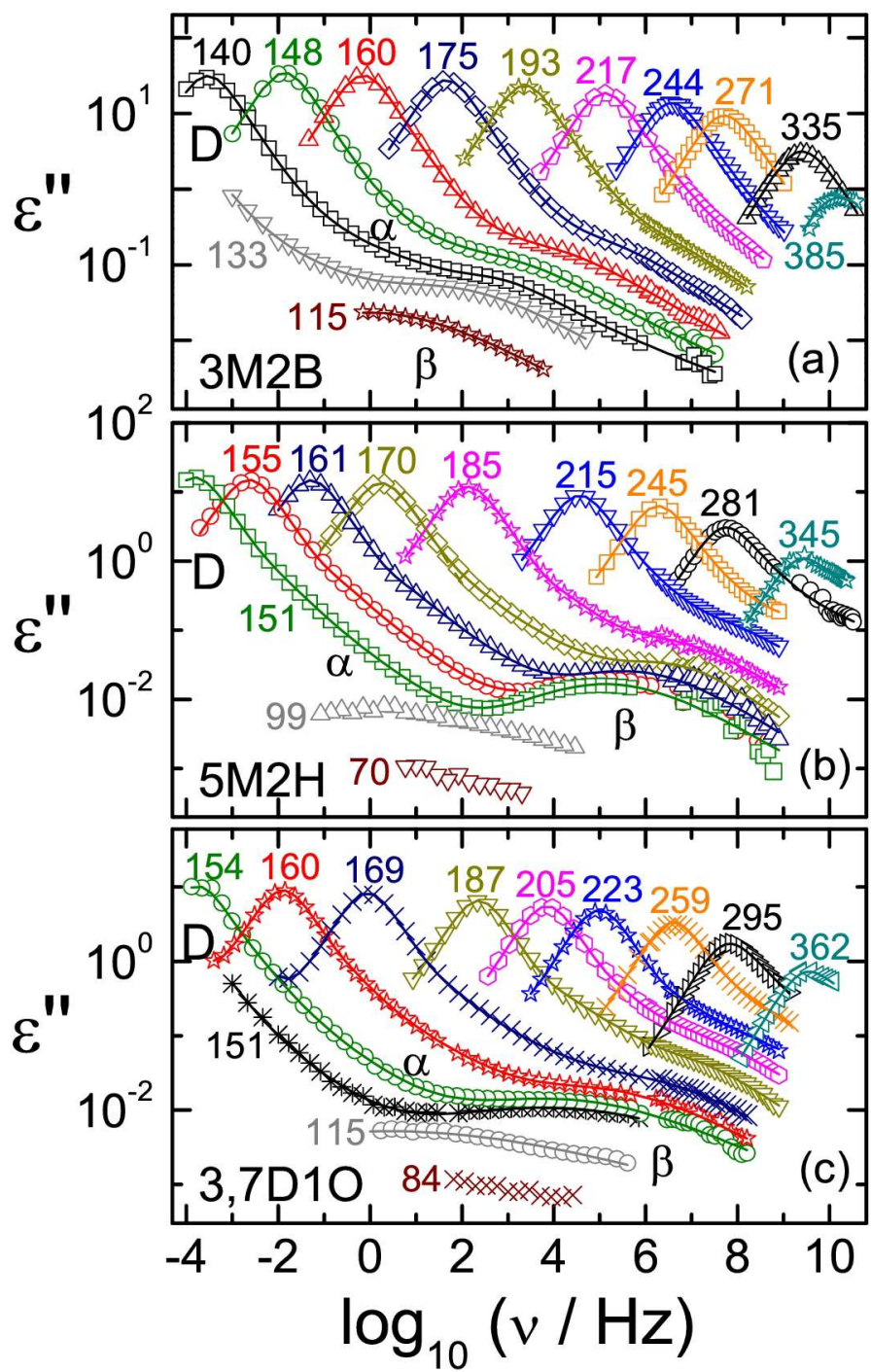

Fig. 2

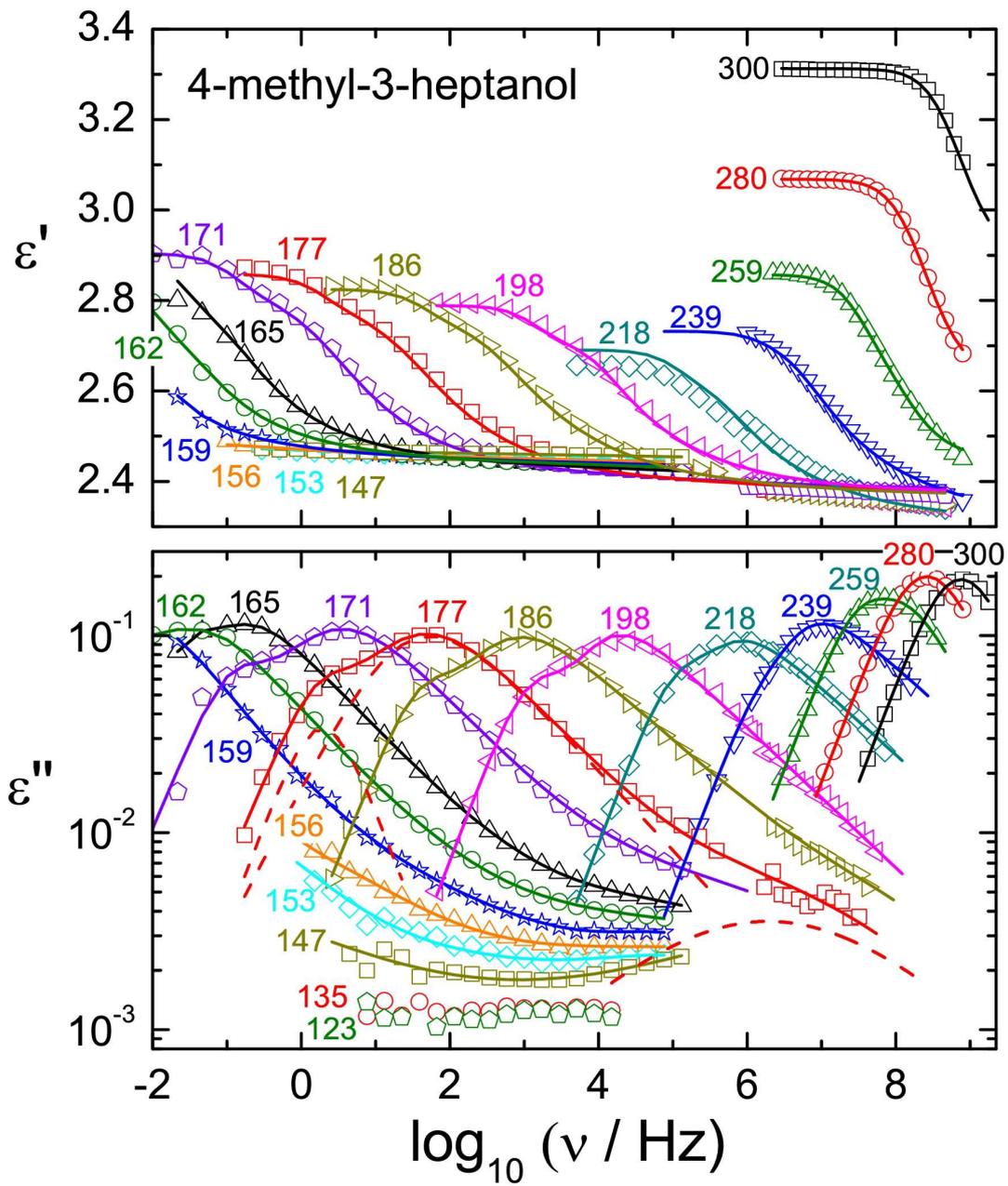

Fig. 3

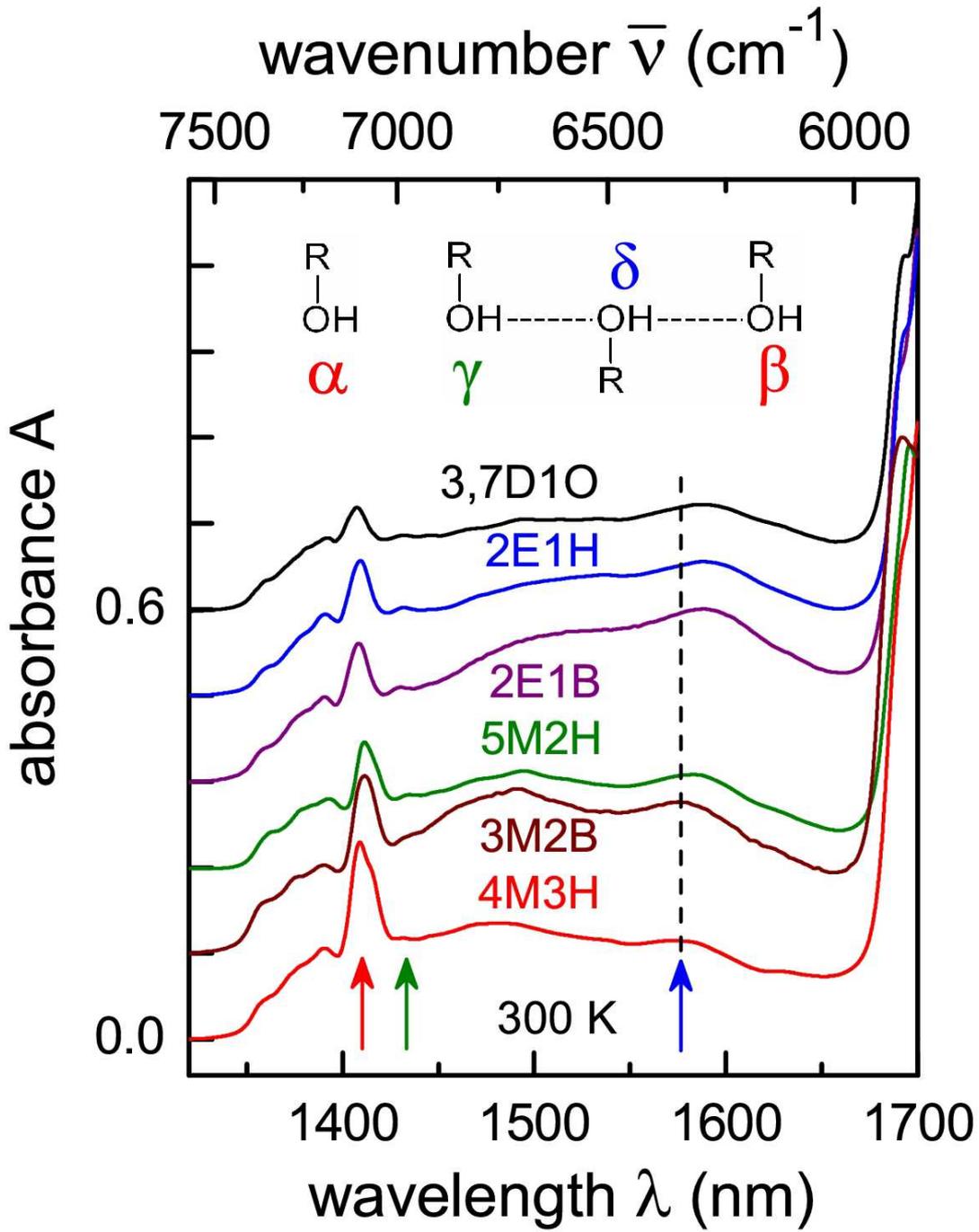

Fig. 4

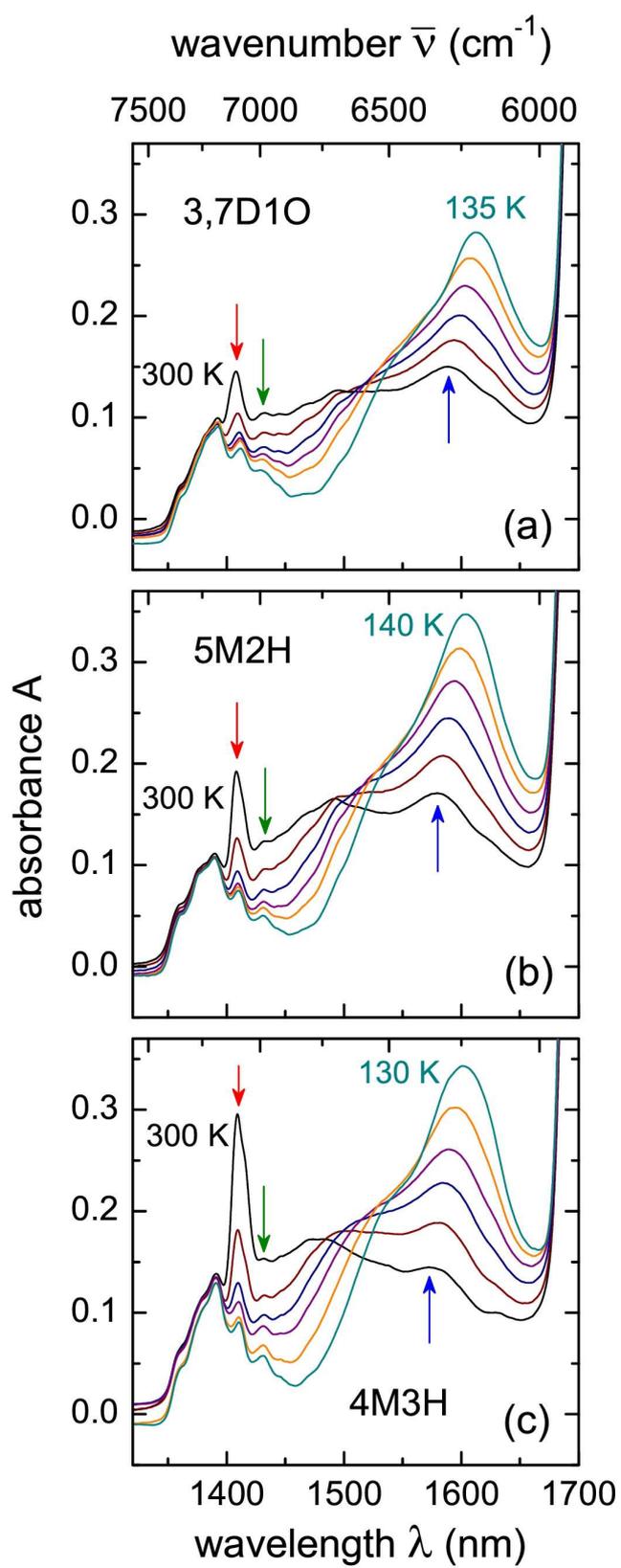

Fig. 5

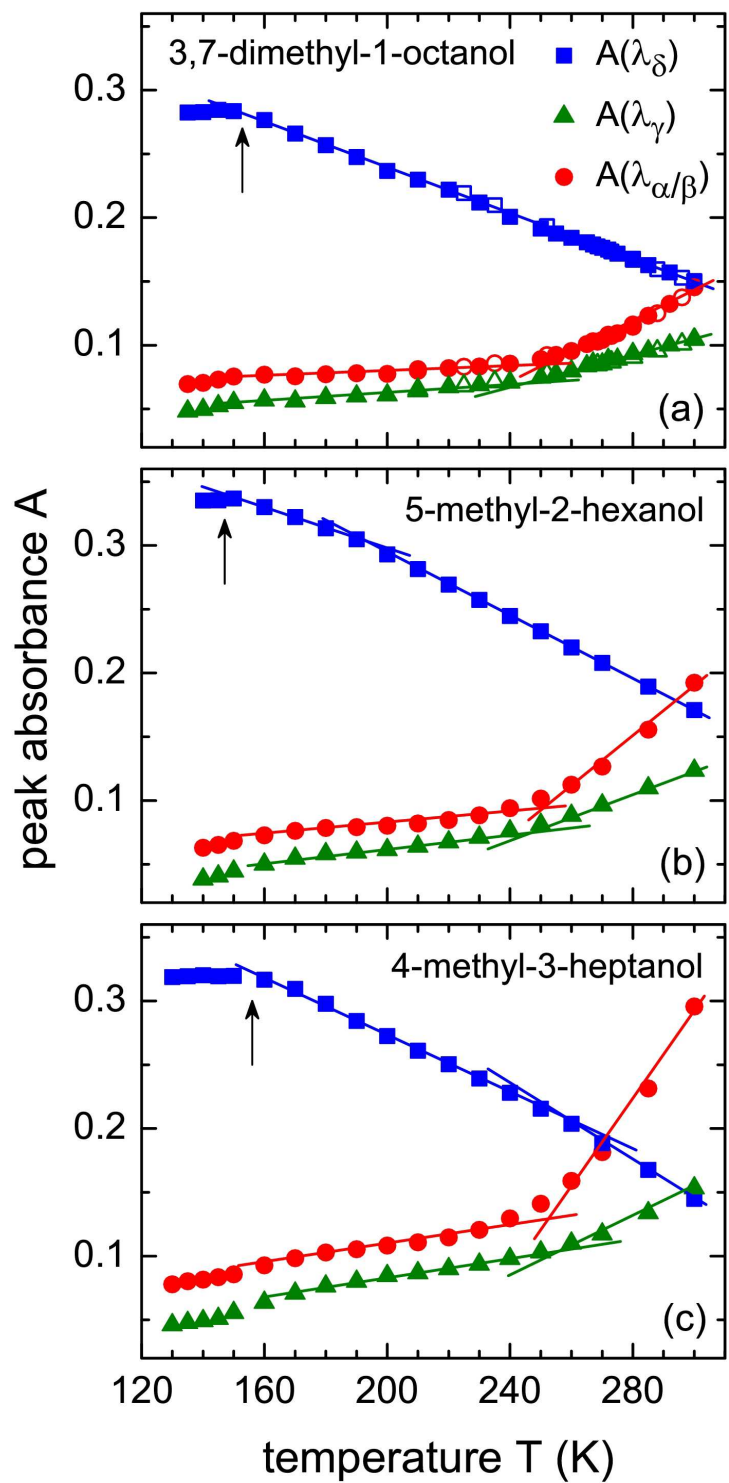

Fig. 6

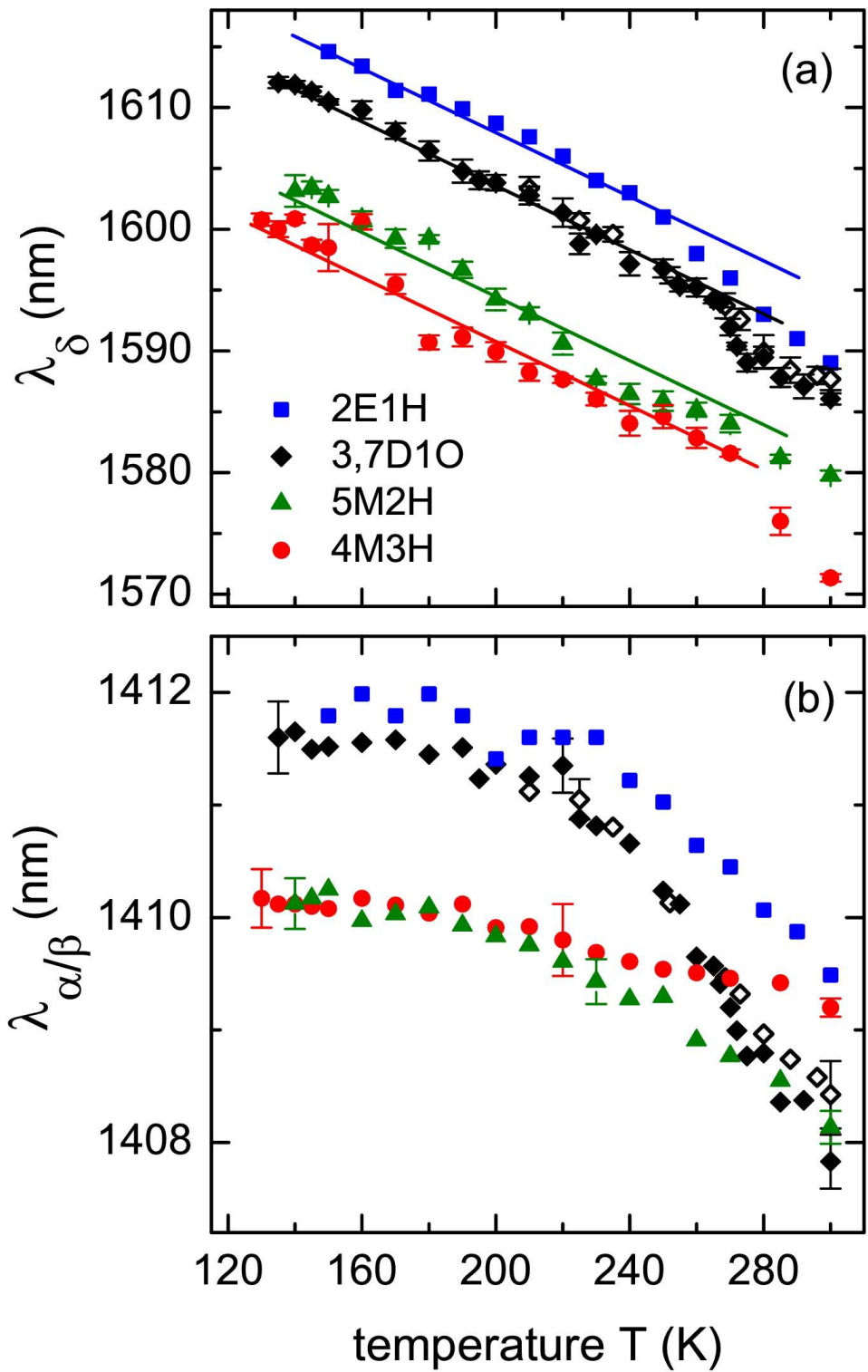

Fig. 7

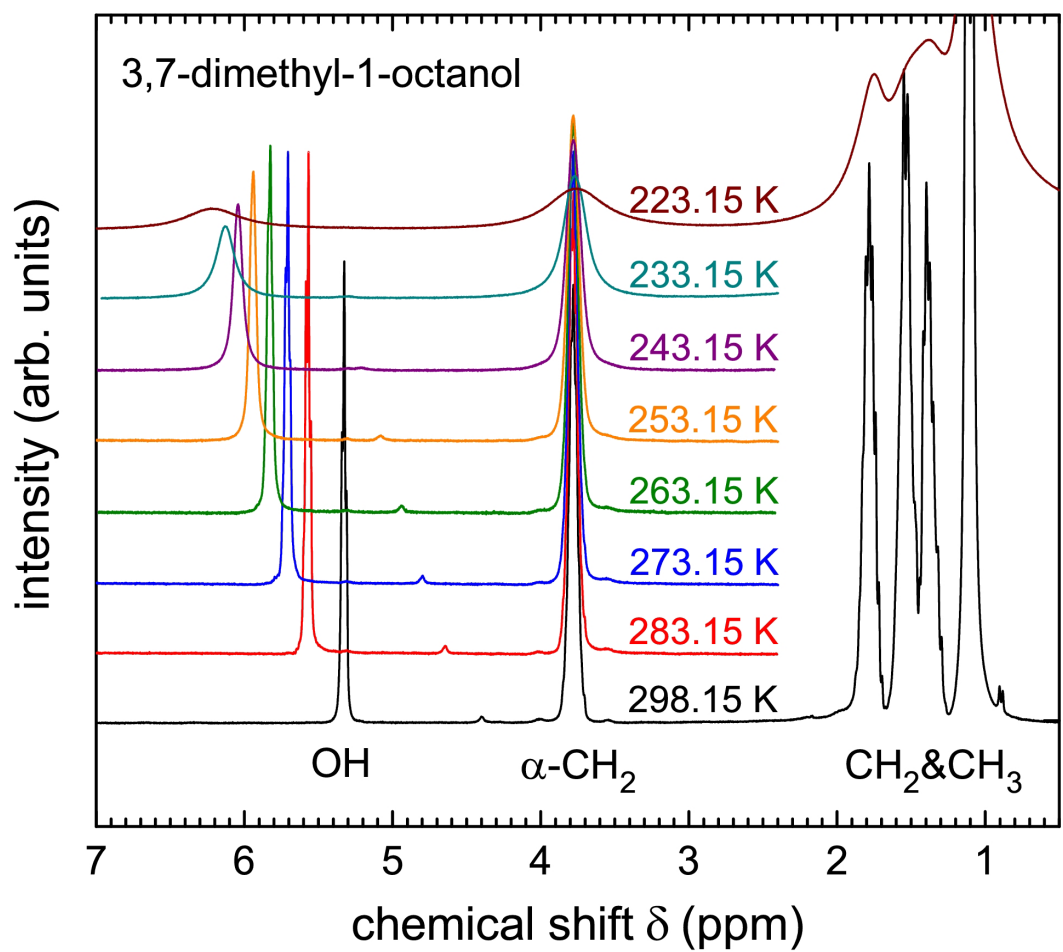

Fig. 8

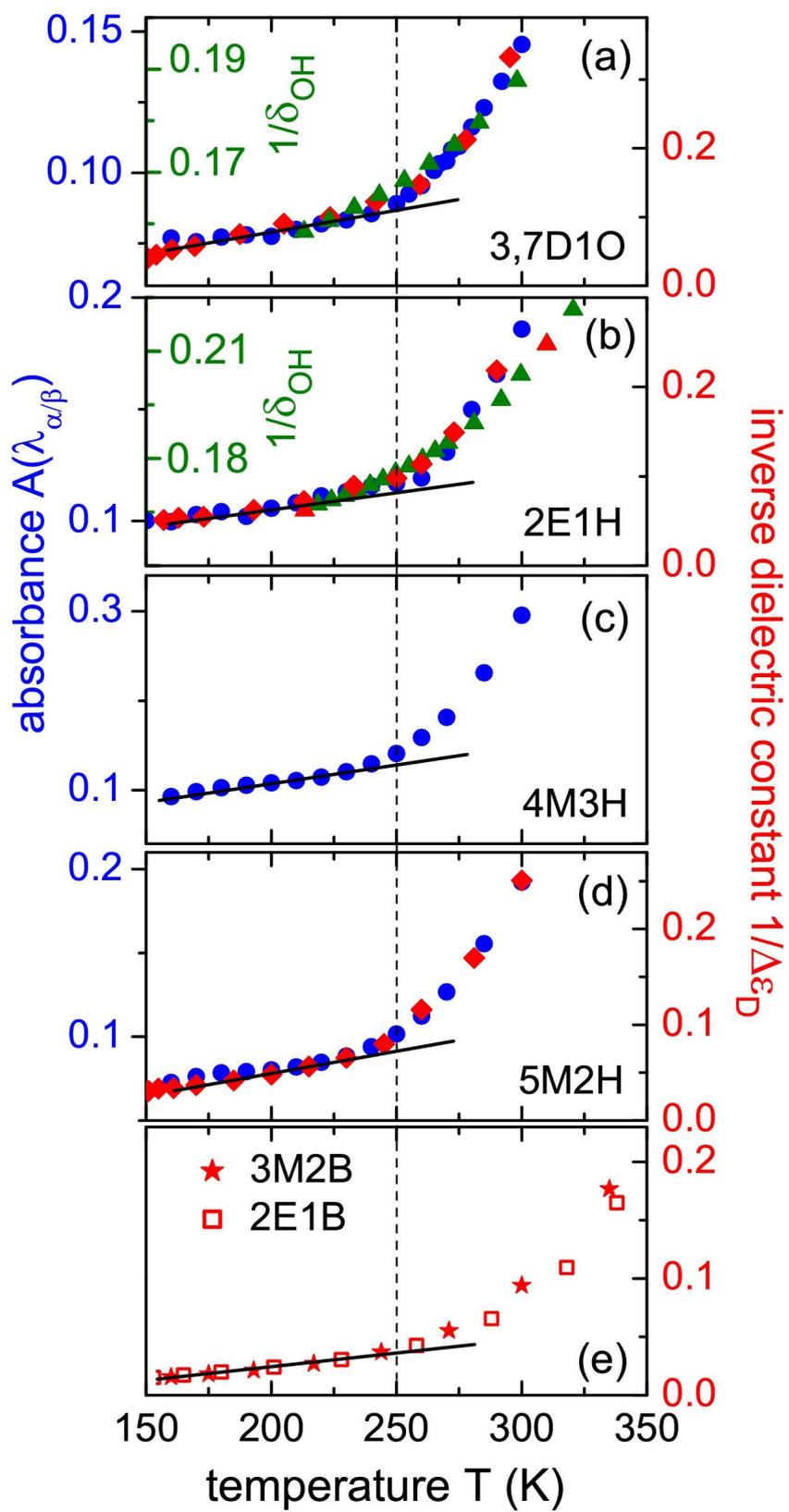

Fig. 9

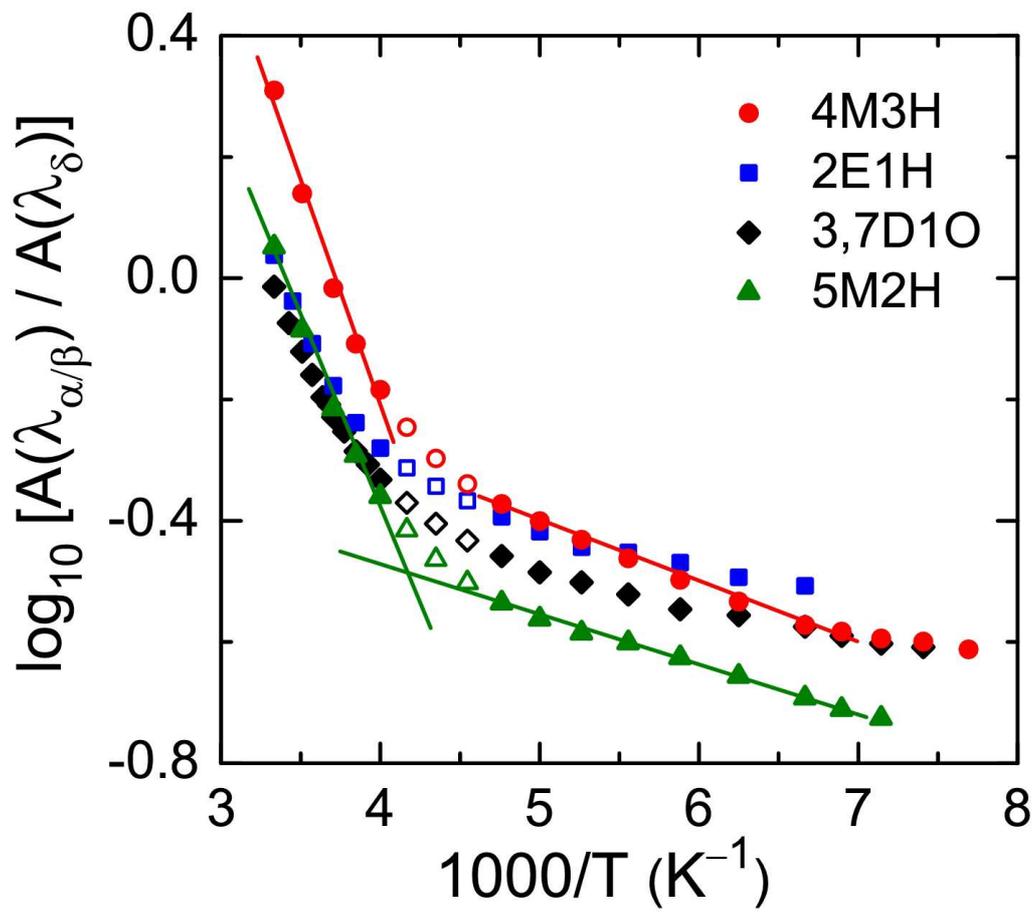

Fig. 10

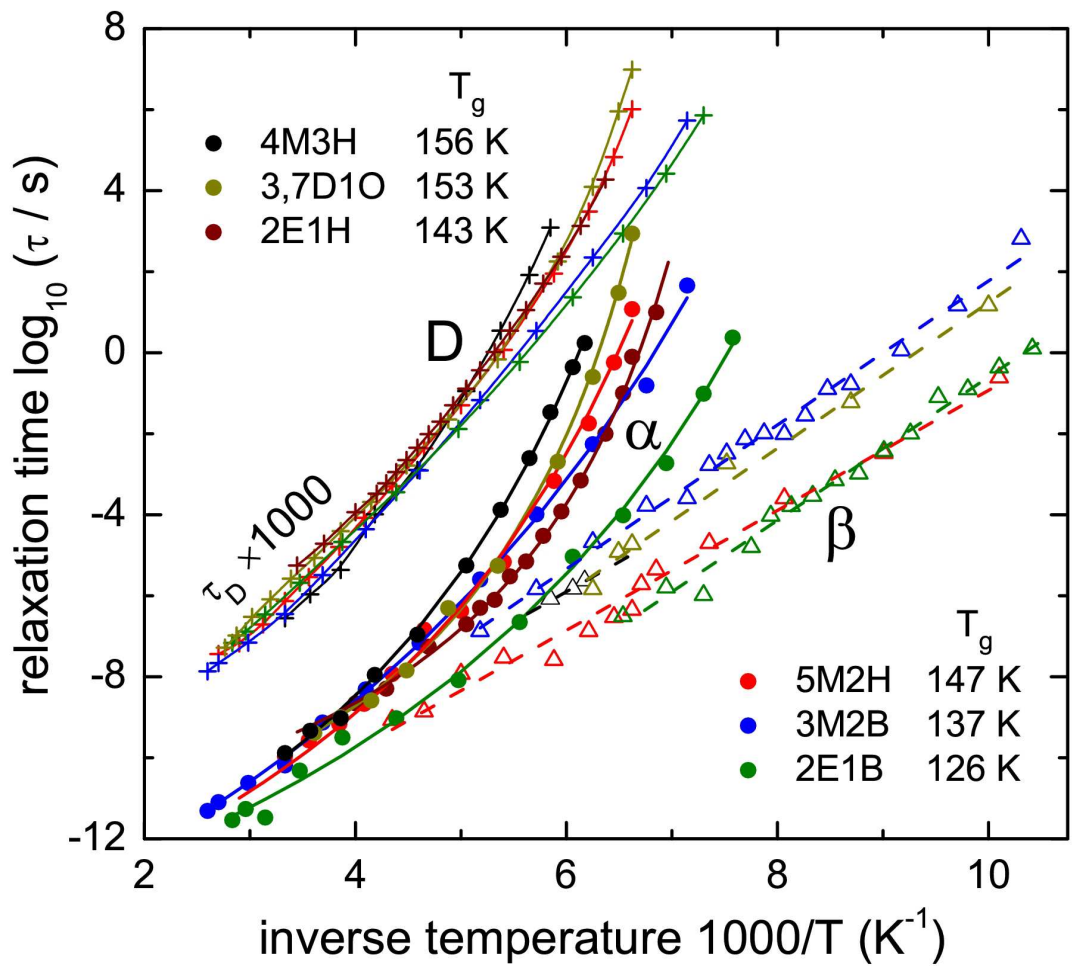

Fig. 11

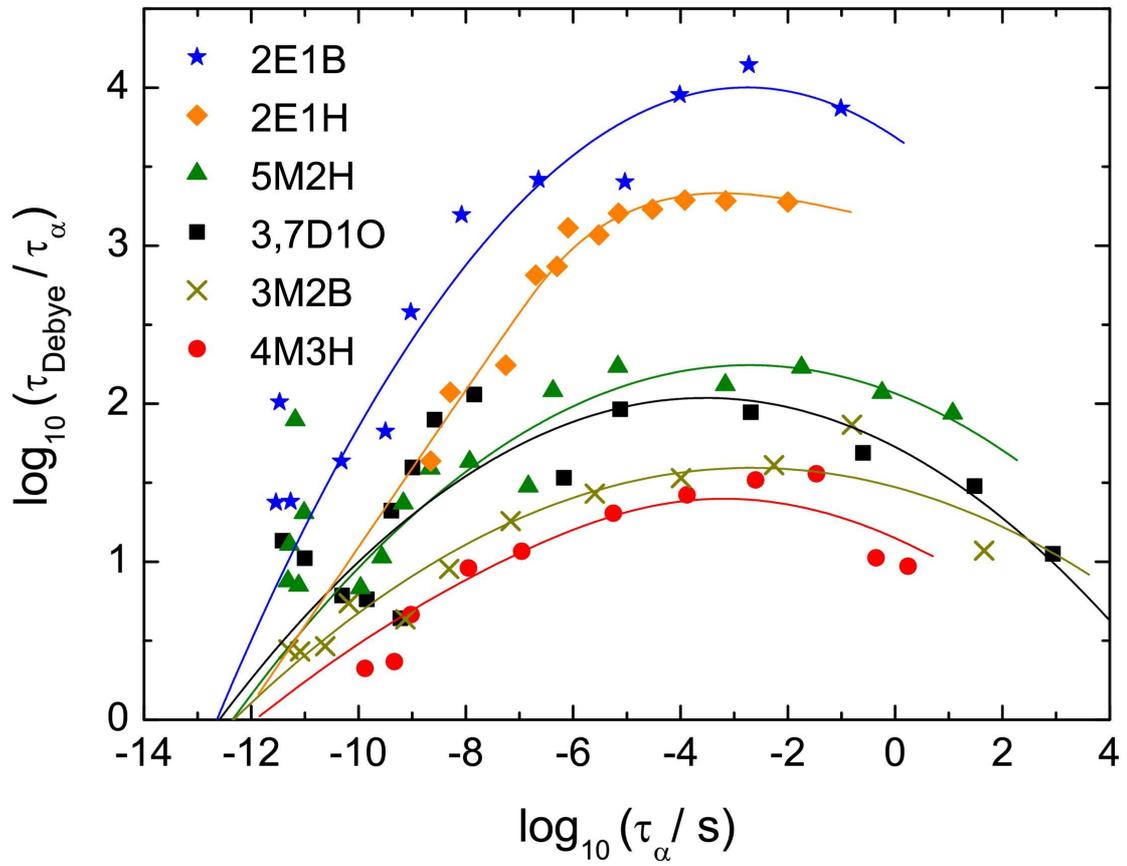

Fig. 12